\documentclass[12pt,english]{article}
\usepackage[T1]{fontenc}
\usepackage[latin9]{inputenc}
\usepackage{setspace}
\makeatletter

\usepackage{ifpdf}

\newif\ifpdf
\ifx\pdfoutput\undefined
  \pdffalse
\else
  \pdfoutput=1
  \pdftrue
\fi

\RequirePackage{xspace} %
\RequirePackage{subfigure} %

\ifpdf

  \RequirePackage[ pdftex, plainpages = false, pdfpagelabels,
                 pdfpagelayout = useoutlines,
                 bookmarks,
                 breaklinks = true,
                 linktocpage,
                 colorlinks = true,
                 linkcolor = blue,
                 urlcolor  = blue,
                 citecolor = blue,
                 anchorcolor = blue,
                 hyperindex = true,
                 hyperfigures
                 ]{hyperref}

\else
  \RequirePackage{color}
  \RequirePackage{colortbl}
   \RequirePackage{array}
  \RequirePackage[dvips]{graphicx}
  \RequirePackage{hyperref}
  \usepackage{rotating}
\fi

\usepackage{makeidx} 
\usepackage{setspace} 
\usepackage{rotating} 
\usepackage{ecltree}
\usepackage{epic}
\usepackage{supertabular}  
\usepackage{color}
\usepackage{exscale}
\usepackage{fontenc}
\usepackage{ifthen}
\usepackage{latexsym}
\usepackage{makeidx}
\usepackage{syntonly}
\usepackage{inputenc}
\usepackage{graphicx}
\usepackage{setspace}
\usepackage{caption2}
\usepackage[english]{babel}
\usepackage[square, comma,numbers,sort&compress]{natbib}
\usepackage{hypernat}
\usepackage{boxedminipage}
\usepackage{framed}
\usepackage{longtable}
\usepackage[all]{hypcap}

\newcommand{\note}[1]{\marginpar[left]{\singlespace \tiny #1}}

\renewcommand{\sectionmark}[1]%
      {\markright{\thesection\ #1}} 

\renewcommand{\note}[1]{}
\renewcommand\@biblabel[1]{$[{#1}]$}

\usepackage[bottom]{footmisc}

\makeatother

\usepackage{babel}
\begin{document}
\begin{doublespace}
\begin{center}
\textbf{\LARGE{}Special Relativity: \\\vspace{0.2cm}Scientific or
Philosophical Theory?}{\Large{} }\vspace{-1.1cm}
\par\end{center}
\end{doublespace}

\begin{center}
Taha Sochi (tsochiATgmailDOTcom)\vspace{-0.4cm}
\par\end{center}

\begin{center}
Independent author, London, UK
\par\end{center}

Abstract: In this article, we argue that the theory of special relativity,
as formulated by Einstein, is a philosophical rather than a scientific
theory. What is scientific and experimentally supported is the formalism
of the relativistic mechanics embedded in the Lorentz transformations
and their direct mathematical, experimental and observational consequences.
This is in parallel with the quantum mechanics where the scientific
content and experimental support of this branch of physics is embedded
in the formalism of quantum mechanics and not in its philosophical
interpretations such as the Copenhagen school or the parallel worlds
explanations. Einstein theory of special relativity gets unduly credit
from the success of the relativistic mechanics of Lorentz transformations.
Hence, all the postulates and consequences of Einstein interpretation
which have no direct experimental or observational support should
be reexamined and the relativistic mechanics of Lorentz transformations
should be treated in education, academia and research in a similar
fashion to that of quantum mechanics.

\begin{flushleft}
\par\end{flushleft}

\section{Introduction}

As it is well-known, the crisis related to a number of critical physical
issues such as the speed of light in free space, the existence of
a luminiferous medium, and the failure of the Maxwell equations to
be form invariant under the Galilean transformations have motivated
the research in the end of the nineteenth century to find a solution
that can address all these inconsistencies and difficulties. Several
renowned physicists of that time, such as Lorentz and Poincare, suggested
scientific solutions and philosophical interpretations to address
these issues. The most prominent of these attempts is the proposal
of Lorentz who suggested certain mathematical transformations of space
coordinates and time that seemed to provide a reasonable solution
to most of those difficulties.

In his attempt to provide a logical interpretation of the Lorentz
transformations while building on the attempts of his predecessors,
Einstein \cite{Einstein1905} adopted the two famous postulates of
special relativity which are the constancy of the speed of light in
free space for all inertial observers and the covariance of the physical
laws in inertial frames of reference. He then derived the Lorentz
transformations from these postulates and extracted a number of consequences
and interpretations. Apart from the formalism of the Lorentz transformations
and their direct mathematical consequences, such as the transformations
of velocity and other physical quantities, Einstein attempt was shrouded
by a number of assumptions and conclusions which are either of philosophical
nature or have no direct experimental evidence. The whole package
of formalism and his interpretation and attachments was then marketed
as the theory of special relativity which, with time, gathered momentum
thanks to the gradual support from experiments and observations and
the endorsement of renowned scientists of that time and hence it gained
general acceptance by the main stream science.

Special relativity of Einstein contains three main elements:

(A) The formalism of Lorentz transformations and their direct mathematical
consequences.

(B) Philosophical issues which are beyond the reach of science and
scientific research such as the abolishment of absolute space and
time.\footnote{It is difficult to give a universally-accepted definition of scientific
and philosophical premises and the difference between them. However,
in this respect we adopt a simple definition to distinguish between
the two, that is a scientific premise is an idea that can be verified
by direct experimental and observational evidence while a philosophical
premise cannot as it has some non-experimentally-verifiable content.
We should also point out that even some premises and theories which
are scientific by nature may not be qualified as such if there is
no unquestionable evidence in their support. }

(C) Scientific issues which are not verified directly by experiment
or observation such as the constancy of the speed of light in free
space for all inertial observers and being the ultimate physical speed
in Nature.

Since the theory of special relativity was accepted as a whole, all
these three categories were accepted as scientific products; moreover
any evidence supporting the relativistic mechanics of Lorentz transformations
was regarded as evidence for the whole package and hence several philosophical
premises and scientific issues which have no direct experimental support
have enjoyed general acceptance thanks to the experimental and observational
support for the formalism of Lorentz transformations. In addition
to all that, the theory of special relativity is treated by many as
the only viable theory and hence the quest for an alternative theory
that can be as good as special relativity or even better has been
stopped and even suppressed as if this theory is the ultimate truth
and the only feasible solution. Different philosophical and scientific
interpretations of the Lorentz transformations which were circulating
in the early days of the above-described historical developments also
ceased to be considered or developed further with the opposition of
any attempt to develop new interpretations to rationalize the formalism
of relativistic mechanics of Lorentz transformations or provide an
alternative formalism, interpretation and rationalization altogether.

At about the same time and in parallel to the development of the relativistic
mechanics, the quantum mechanics was also under development. Similar
to the relativistic mechanics, the quantum mechanics was made of two
main parts:\footnote{For the sake of simplicity in presentation, we can classify the above
(C) category as philosophical although the issues in this category
are scientific by nature.} a formal mathematical part and a philosophical interpretive part
where these two parts were developing side by side. While the quantum
mechanics kept the separation between these two parts and have generally
considered the experimental and observational evidence as support
to the formalism only and not to any particular interpretation, the
relativistic mechanics did not develop in this way and hence the two
parts were mixed in a single entity in the form of special relativity
thanks to the dominance of Einstein reputation and his many supporters,
as indicated above.

\section{Examples of Challenges to Special Relativity}

Special relativity of Einstein has been challenged by many scientists
and philosophers in the early years of its release. Nowadays, there
are still challengers and opponents to special relativity but this
opposition is less fierce and sometimes it is shy and hidden due to
the coordinated and well organized efforts to suppress any opposition.
We believe that not all criticisms and concerns about the logical
foundations and scientific premises of special relativity have been
addressed properly. This does not mean that we endorse these criticisms
and concerns, but we think these should have been and must be addressed
properly. A few examples of these legitimate criticisms and concerns
are:

(A) The twin paradox where different conflicting answers have been
given to address this paradox such as calling for general relativity
or explaining the difference by the acceleration and deceleration
of the traveling twin. Most of these answers are arbitrary, superficial
and lack depth and content. We are not here to discuss and analyze
these answers so we refer the reader to the vast literature about
this paradox hoping we may come back to this matter in the future.
However, even if we accepted all the answers provided to address the
original version of the twin paradox, no convincing answer has been
given to some of the revised versions of this paradox. In particular,
we should mention Dingle discussion \cite{DingleBook1972} of this
issue among other issues and the rejection which he faced.

(B) The claims made by a number of scientists about the variation
of the speed of light or exceeding this speed (i.e. the value $c\simeq3\times10^{8}$
m/s) in some astronomical observations and experimental proceedings
where this speed is supposed to be the ultimate speed for any physical
movement according to special relativity. In particular, we refer
to several claims from respected scientists questioning the constancy
of the speed of light and also to the recent experiment in the Large
Hadron Collider (LHC) where the initial claim of exceeding the speed
of light vanished suspiciously by a mysterious technical fault in
the electronics and connections.

We repeat that this is just a sample of the logical and scientific
inconsistencies and controversies which have been or can be claimed
to be contained in special relativity. As we indicated above, it is
not necessary that we agree on the validity of these criticisms; what
these criticisms highlight is that the philosophical and logical foundations
and interpretations and some scientific attachments of the relativistic
mechanics of Lorentz transformations as embedded in special relativity
require reexamination since there are challenges and claims which
are not addressed properly, and from what we are aware of most of
these rejections are based directly or indirectly on the experimental
and observational support which the relativistic mechanics has enjoyed
where this support, in our view, can be used as evidence for the formalism
of the relativistic mechanics of Lorentz transformations but not for
Einstein interpretations and attachments.

\section{Importance of Philosophical Interpretations of Scientific Theories}

Although there may be a general tendency to restrict science to its
experimental and observational domain and separate it from any association
with philosophical issues, the reality is that no science is completely
free from such philosophical issues at least at the level of postulates
and primary premises which the science rests upon especially for the
theoretically-oriented natural sciences. Moreover, the philosophical
ideas and interpretations which are embedded in science usually provide
an insight in the formalism and why Nature behaves in a certain way
by trying to rationalize its behavior. After all, science is an attempt
to understand the physical world and make sense of it, and this may
explain why science does not stop at the formal rules and mathematical
formalism even if these work perfectly but it tries to provide explanatory
statements of descriptive and qualitative nature to make sense and
rationalize the rules and formalism. Adding to all this, the philosophical
ideas, which may stay in the background of the scientific theory,
provide an incentive and steering force which points to the direction
that should be followed in the future development and hence they act
like a dynamo and compass for scientific research. Therefore, the
importance of the philosophical component of any scientific theory
should not be underestimated where this underestimation can lead to
a rejection of the call for reexamination of certain philosophical
issues justifying this by an excuse that as long as the formalism
is working we can keep the philosophical framework of the theory even
if it is controversial or imperfect or even wrong.

\section{Philosophical Issues}

As indicated above, and apart from the unverified scientific elements,
special relativity contains philosophical elements which are beyond
the realm of science and hence these issues should be clearly distinguished
from the formalism of relativistic mechanics and should be deprived
from the automatic endorsement which they enjoyed so far from any
real or alleged success of the relativistic mechanics. A prominent
example of these issues is the claim that special relativity has abolished
absolute space and time forever. Views which sound like ``Einstein
changed our understanding of the world forever by abolishing Newtonian
space and time'' are very common in the literature of special relativity.
Apart from being implicit statements of reaching the final truth,
no one can abolish such fundamental and valuable ideas like absolute
space and absolute time, neither Einstein nor anybody else. Assuming
that these fundamental concepts are not built into the roots and fabric
of our intellectual structure and mental blueprint like the basic
rules of logic, they are at least very useful conceptual tools which
have been and will remain available and useful for building philosophical
and scientific theories that can provide better understanding and
superior physical and conceptual adaptation to the outside world.
If Einstein chose to abolish these concepts from his view of the world,
as represented by his theory, he is free to do so, but neither he
nor anybody else can abolish these fundamental ideas from the view
of the world of other thinkers or deprive them from the legitimacy
to exist and be used in science or philosophy or any other discipline
if they are found to be useful, convenient and illuminating.

As an indication to how fundamental these concepts are and as a sign
to the failure of this alleged abolishment, there are many places
in special relativity where one feels that these concepts are still
there, at least in the background, and they are still needed to present,
formulate and visualize the settings of special relativity itself
and its physical environment despite the apparent denial of their
existence. This will be automatically rejected, as usual, explaining
this by our habits, cultural heritage and likewise but, apart from
the fact that all these excuses can be challenged, this is just another
indication of how fundamental these concepts are when they propagate
even to our habits and cultural heritage. Anyway, any answer to this
concern will certainly fail to put an end to these controversies;
what is important of all these is that Einstein philosophical views
are not final or universal and they could be and should be challenged
at least for the sake of motivating further progress in science and
human thinking in general.

\section{Comparison between Relativistic Mechanics and Quantum Mechanics}

Returning to the above comparison between the development of relativistic
mechanics and quantum mechanics, despite the fact that Lorentz transformations
have been interpreted differently, at least in the early days of relativistic
mechanics, or can in principle be interpreted so, we see the dominance
of Einstein interpretation. As a consequence, Einstein views of relativistic
mechanics are presented in academia and education and employed in
scientific research as the sole interpretation of the Lorentz transformations.
The theory of Einstein contains several philosophical issues and scientific
aspects which cannot be proved or have not been proved directly by
experiment or observation. The total acceptance and complete adoption
of Einstein views made the relativistic mechanics the mechanics of
Einstein theory as if it is the only possible interpretation and made
all the philosophical and scientific issues, whether they can be proved
or not or have been proved or not, scientific facts which automatically
get credit and support from any genuine or alleged success of the
relativistic mechanics of Lorentz transformations.

Unlike relativistic mechanics, quantum mechanics is presented and
adopted in its bare formalism as the scientific theory while its different
interpretations are appended as possible explanations to rationalize
the stated formalism. We believe that, for the sake of equality and
scientific impartiality, if not for any other reason, relativistic
mechanics should be treated like quantum mechanics, and hence relativistic
mechanics should be presented in a more formal and scientific fashion
through its bare formalism which can then be followed by Einstein
interpretation and his attachments, or any other interpretation and
attachment, if this is needed at all.

\section{Relativistic Mechanics in Education, Academia and Research}

Regarding the presentation and treatment of relativistic mechanics
and special relativity in education, academia and scientific research
we have a number of points:

(A) We call for a total revision of the structure of relativistic
mechanics so that it is presented and treated like quantum mechanics,
that is the bare formalism of the relativistic mechanics should be
stated as the real scientific theory, while any philosophical or logical
interpretation (prior or post) should be appended, if necessary, and
labeled as such. Einstein interpretation should then be treated like
any other interpretation where the audience will have the freedom
to choose and embrace any interpretation if they decided to do so
or even to develop their own interpretation. This will not affect
the efficiency and productivity of the educational, academic and scientific
process; in contrast it should improve understanding and creativity.
Quantum mechanics is certainly more subtle and difficult to understand
and harder to make sense of than special relativity; however, there
is no known failure in education, academia or research due to lack
of foresight. Anyone can learn and apply the rules of quantum mechanics
with no difficulties even if he is not aware at all of any interpretation
or school of thought to rationalize these rules. A major advantage
for adopting the formalism of relativistic mechanics of Lorentz transformations
instead of special relativity is that the possibilities will be open
for investigations in directions forbidden by special relativity such
as the possibility of considering speeds exceeding the speed of light,
at least in some theoretical speculations, to prevent blocking possible
venues for scientific progress. The fanatic adoption of special relativity
which includes some questionable taboos may hinder the scientific
progress in some aspects and directions of research.

(B) We also call for abandoning the personalized presentation of Einstein
and his views in education, academia and research and the deliberate
and non-deliberate endorsement not only of his views but also of his
personal credential. We call for abandoning these types of exaggeration
which is shrouded with mystic affection and religious devotion. This
type of presentation may seem enthusiastic, entertaining and motivating
but it certainly leads to worshiping of individuals, killing the spirit
of creativity and giving a wrong message that certain individuals
are super-human and certain theories are final and cannot be challenged
or replaced. This sort of presentation, which is very common in the
circulating literature about Einstein and his theories, is not only
against the spirit of science and a divergence towards a religious-like
loyalty and devotion but it also inflicts costly damage to the educational,
academic and scientific process.

(C) There are many instances where the challenge to special relativity,
and Einstein views in general, is depicted implicitly or explicitly
as motivated by personal grudge or envy to Einstein and his legacy
or a conspiracy to defame him and devalue his legacy. Although there
may be historical reasons to believe that some of the early opposition
to Einstein views was motivated by such factors, the contemporary
opposition is generally different in motives and objectives. It should
be understood that challenging Einstein is like challenging any other
scientist and the majority of opponents these days are not motivated
by hate or jealousy or any other emotion or conspiracy, but different
people have different views and something that looks obvious and correct
to an individual can look absurd and wrong to another. This sort of
personalizing the debates and challenges, which takes special importance
in the debates related to special relativity, should be stopped and
any opposition to this theory or any other theory should be dealt
with scientifically and professionally without involving any kind
of accusation or demoralization which are professionally and morally
wrong.

(D) There are many direct and indirect suggestions in the literature
of special relativity that ``understanding'' (which usually means
accepting) special relativity is an intelligence test. In fact special
relativity is one of the simplest theories in physics, all is needed
to understand this theory is a modest imagination with a little arithmetic
and basic algebra plus the Pythagoras theorem; all of these are taught
at the final years of primary schools or the first years of secondary
schools. If one can understand statistical mechanics and quantum mechanics
he should be able to understand special relativity. What some people
find difficult to ``understand'' (or rather digest) about special
relativity is the consistency of its logical and philosophical structure
as these people feel there are potential inconsistencies in this theory,
and that is why they fail to ``understand''. Anyway, no theory should
be put on the list of IQ test and hence these attempts to degrade
any opposition or questioning of the validity of special relativity
should disappear from education, academia and scientific research.
In particular, labeling these attempts as ``crackpot'' which is
common even in some respected scientific circles should stop at least
for the sake of courtesy and politeness if not for the sake of free
thinking and scientific spirit.

(E) Assuming that all the theories and views of Einstein are completely
right, the search for alternative theories should never stop because
our search for alternative theories should not be restricted to where
we believe that our theories may be wrong or incomplete but even where
we believe that our theories are totally perfect. Our search for the
better, and even the equal, should continue not only for the purpose
of enjoying the leisure of thinking and creating models and ideas
but also for many practical reasons. What looks completely right today
may not look so tomorrow. Having alternative theories will always
make us richer and more comfortable in dealing with any future challenges
whether theoretical or practical. In fact there are many examples
in science and mathematics where we have more than one correct theory
or method, one of which is better to use in one context while another
is better in another context. So even if special relativity is completely
correct in our view, a critical examination of relativistic mechanics
and Einstein views and the search for alternative theories must continue.
Of course this critical examination and search for alternatives should
apply to all branches, theories and views in science and is not limited
to relativistic mechanics and Einstein views.

\section{Issues to be Reexamined}

The following premises, which are largely regarded as direct or indirect
implications and consequences of special relativity, should be considered
as potential candidates for reexamination either because they are
philosophical by nature, rather than scientific, or because they have
no sufficient direct evidence from experiment and observation:

(a) The abolishment of absolute space and time.

(b) The abolishment of the independence of time from space.

(c) The relativity of simultaneity.

(d) The interpretation of time dilation effect, at least in some of
its forms.

(e) The interpretation of length contraction effect, at least in some
of its forms.

(f) The constancy of the speed of light in all inertial frames.

(g) The imposed limit of $c$ as the ultimate speed for any physical
object.

(h) The logical foundations of the theory and if it is self-consistent
and consistent with the rules of logic in general or not.

The call for reexamination of these issues does not mean that the
above issues are already classified; all we say is that these issues
require reexamination so that any acceptance or rejection of one of
these issues should be based either on a well established philosophical
and logical stance or on direct experimental or observational evidence.
As stated above, some of the above issues, such as the abolishment
of absolute space and time, are philosophical by nature and hence
they are beyond the domain of scientific verification. Any theory
has full right to adopt its own concepts, definitions, conventions
and philosophical framework as long as it is self-consistent and does
not lead to illogical consequences and absurdities or conflicts with
experiment and observation.

\section{Conclusions and Final Thoughts}

We summarize the main issues discussed in the present article in the
following bullet points:

$\bullet$ We call for a separation between the formalism of relativistic
mechanics of Lorentz transformations from any unverifiable philosophical
issues or unverified scientific attachments. We also call for thorough
tests for the logical structure of special relativity to see if it
is consistent or not as claimed by some.

$\bullet$ With regard to the abolishment of absolute space and time,
apart from the above rejection of this alleged abolishment, there
is no sensible meaning of this abolishment which is largely claimed
by the followers of special relativity. Concepts like ``absolute
space'' and ``absolute time'' are no more than conceptual tools
that can be used in any physical or philosophical theory if they proved
to be practically useful or can provide a better insight and sense
of direction in our physical and conceptual adaptation to the physical
world. In fact, denying these concepts and replacing them with alternative
concepts like ``relative space'' and ``relative time'' is based
on a more extreme belief in absoluteness and objectivity of the physical
reality as depicted by our theories and our conceptual models than
the ones that are supposed to be abolished. This feeling of ``absolute
relativity'', which is based on a strong belief in the objectivity
of the relativistic definition of space and time, should be replaced
by a more modest stand about the reality of these relativistic concepts.

$\bullet$ Even if special relativity appears to us as the only viable
interpretation of Lorentz transformations or the most endorsed interpretation
by logic and science, the door should be kept open for potential interpretations
that may emerge in the future at least out of respect of the spirit
of science and for the sake of our continuous search for a better
understanding to the world. The fanatic rejection of any examination
of this theory will block any attempt to advance the science in this
very important field of physics and remind us of the dogmatic embracement
of Aristotle and his science and philosophy which resulted in hindering
the scientific progress and free thinking for many centuries.

$\bullet$ All attempts to degrade and demoralize critics of scientific
theories should be stopped especially those attempts which violate
the rules of courtesy and civility like labeling certain types of
criticism as crackpot. No theory should be treated as the final word
or embraced with the zeal of a religious belief.

$\bullet$ It should always be kept in mind that we did not reach,
and will never reach, the ultimate ``truth'' neither in relativistic
mechanics or in any other field of knowledge. We should not repeat
the mistakes of our ancestors, who at many stages of history believed
that they reached the ultimate truth and hence they kept consuming
ideas and practices which proved to be imperfect, to say the least,
and can be improved. Knowledge, and science in particular, is based
on a continuous search for the ``truth'' and relentless attempts
to improve our understanding and adaptation to the outside world.
Even if the existing theories are completely correct, we should still
keep searching for alternative theories and expand and improve the
existing ones. A correct theory may be replaced by another correct
theory which is simpler to understand and use for instance. We should
not allow fundamentalist views to exist and flourish in science as
long as we believe that science should keep growing and improving.

$\bullet$ The word ``forever'' should be removed from the vocabulary
of science in general and relativistic mechanics in particular where
this word occurs very often. There is no ``forever'' in science
since it is a continuously developing and improving process. Science
becomes forever only when it is dead and mummified. Some people seem
to think that a day will come when science becomes complete and we
have the final version of the ``truth'' and hence they are looking
for that day where we all retire from doing science and all we do
then is to keep consuming perfect theories and practices which were
developed by our ancestors. This view, which sounds like a fundamentalist
religious view, is against the law of evolution which is one of the
most fundamental laws of Nature, at least from our current viewpoint.
In fact even the constancy of the laws of Nature in time and space
should not be taken as a well-established and undisputed fact. So,
if Nature itself is potentially changing its rules in space and time
then there will be no point in looking for the final version of these
rules. Anyway, science is no more than conceptual models and not perfect
imprints and reflections of the rules of Nature, therefore even if
these rules are fixed spatially and temporally there is always space
for modification, perfection and alternatives in our models and views
of the world. There is an element of creativity in science which makes
it a mix of objective impressions with subjective added values.

\pagebreak{}

\phantomsection
\addcontentsline{toc}{section}{References}
\bibliographystyle{unsrt}

\vspace{0.5cm}

\end{document}